# Microtron for Smog Particles Photo Ionization

*S. N. Dolya*

*Joint Institute for Nuclear Research, Joliot - Curie 6, Dubna, Russia, 141980*

**Abstract**

The article discusses a possibility of removing smog particles from a boiler smoke. To do this, the boiler smoke is passed through a flow of gamma radiation, formed by interaction of the microtron beam with a heavy target. The energy of the microtron electrons $\epsilon$ = 25 MeV, the beam current I = 100 µA. Smog particles are ionized with gamma radiation and then sat down on the plates of the electrostatic filter. The height of the filter plates is 1 m, the electric field between the plates E = 1 kV / cm. The smog particles on the plates should be removed regularly to a specialized dust collector.

**1. Introduction**

The main source of smog is boilers using brown coal. Considering that the calorific value of the brown coal is 6.7 Mcal / kg, we find that one ton of this coal releases ~ $3*10^{10}$ J / ton. Taking into account that there are approximately about $10^5$ seconds per day we find that the power of the boiler burning one ton of the brown coal per day will be: P = $3 * 10^{10}/3*10^5 = 10^5$ W. The boiler with power P = 30 MW should burn 100 tons of the brown coal per day.

We assume that one percent of the mass of coal is spent to form smog - fine particles with a diameter of less than 10 µ.

We suppose that the concentration of these particles in the air is 100 µg / $m^3$. This means that the particles will fill the volume around the boiler with the size of about 10 km * 10 km * 100 m. Indeed, this volume contains $10^{10}$ $m^3$ of air and if in every cubic meter of the air there is 100 µg / $m^3$ of the smog particles the total mass of the particles will be equal to 1 ton. Remember that this is the number of smog particles which the boiler produces per day.

Below we will consider a possibility of photo ionization of these particles with gamma radiation produced by a small microtron located on the boiler tube. The removal of the smog particles from the boiler smoke is carried out by means of the electrostatic filter.

## 2. Microtron

Microtron is a compact electron accelerator having a variable fold of acceleration. We assume that the final energy of the electrons in the microtron is equal to Є = 25 MeV, the electron current is taken to be equal to I = 100 μA. Thus, the power of the microtron beam will be equal to P = 2.5 kW.

The frequency of the electron rotation in the magnetic field of the microtron is as follows:

$$\omega_H = eH/mc\gamma, \qquad (1)$$

where e = 5 * $10^{-10}$ which is - the magnitude of the electron charge in the CGS, H – magnetic field, m = $10^{-27}$ g - electron mass, c = 3*$10^{10}$ cm / s - the velocity of light in vacuum, γ = Є / $mc^2$ is the relativistic factor, i.e., the ratio of the electron energy to the rest energy of the electron. During acceleration γ varies from the initial value of γ = 1 to the finite value γ = Є / $mc^2$ ≈ 50.

The microtron has a peculiarity: the product of the magnetic field H multiplied by the wavelength λ of RF electric field, which performs acceleration, must be less than:

$$H*\lambda < 10 \text{ kGs*cm}. \qquad (2)$$

Let us take the wavelength of the acceleration λ = 2.5 cm, which corresponds to the acceleration wavelength in the compact linear collider CLIC at CERN [1]. Selecting of this wavelength acceleration means that the magnetic field of the microtron, retaining the accelerated electrons on circular orbits, must be less than H <4 kGs.

Now we take the magnetic field of the microtron equal to: H = 3 kGs. The frequency of the electron rotation with finite energy according to the formula (1) is then equal to: $\omega_H$ = eH / mcγ = 5* $10^{-10}$ * 3 * $10^3$ / ($10^{-27}$ *3* $10^{10}$ * 50) = =$10^9$. The radius of rotation of the electron having energy Є = 25 MeV in this magnetic field will be equal to:

$$r = c/\omega_H = 30 \text{ cm}. \qquad (3)$$

Thus, this microtron will have a diameter of the microtron magnet pole d = 2r = 60 cm, and it is compact enough to be placed on the boiler tube. The magnetic field strength H = 3 kGs may be formed with permanent magnets consisting of NdFeB, and having the residual induction of the order of 10 kGs.

**3. Gamma radiation of the microtron on a heavy target**

Note some features of the emission of gamma quanta while interacting of the beam of the microtron with a heavy target.

Gamma radiation has a narrow direction. The value of angle θ of the radiation direction is equal to: θ = 1 / γ. This means that it is necessary to use specialized radiation scatters to irradiate the smoke in the tube.

The intensity of radiation decreases inversely proportional to the energy of gamma quanta $E_υ$ . The more is the energy of gamma quanta, the less number of them is emitted. At the current of the electron beam equal to 100 μA and the target thickness of 0.01 of the radiation length, the intensity of the photon radiation is approximately equal to $6*10^{12}$ / $E_υ$ photons per MeV [2]. The radiation length is the length of material where the intensity of gamma radiation decreases by e times, where e is the base of the natural logarithm. For lead the radiation length is 5 mm, for tungsten - 3.5 mm. In this case, in the target having thickness of 0.01 of the radiation length, the microtron beam will produce $6 * 10^{15}$ photons with the energy $E_υ$ = 25 keV.

**4. Photo ionization of the smog particle**

We estimate the probability of photo ionization $p_υ$ of smog particles per incident photon. The corresponding relation may be written as follows:

$$p_υ = σ_υ n l_e, \qquad (4)$$

where $σ_υ$ – photo ionization cross section of the molecules contained in the particles of smog, n ~ $10^{22}$ mol / $cm^3$ - the density of matter in the smog particle, $l_e$ is the distance of the electron free running formed in the result of photo ionization of in the substance of smog particles.

The cross section of photo ionization of molecules $σ_υ$ decreases with increasing the energy of gamma quanta. On the other hand, at increasing the energy of gamma quanta the energy of a free electron grows. This growth is accompanied with the increase of the length of the free electron running in the substance of the smog particles. Therefore, the product $σ_υ * l_e$ remains approximately constant regardless of the energy of the incident gamma photon.

The literature [3], p. 961 gives as follows: for the gamma quantum energy of $E_υ$ = 25 keV the photo ionization cross section is equal to $σ_υ$ = 6 * $10^{-24}$ $cm^2$, the length of the free electron running with the energy of 25 keV in the smog particle substance [3], p. 957, is equal to $l_e$= 3 μ. Then the probability of photo ionization of smog particles by photons is approximately equal to $p_υ = σ_υ n l_e$ =6 * $10^{-24}$ * $10^{22}$ * 3 * $10^{-4}$ = 2 * $10^{-5}$.

We estimate the bulk density of photons per cubic centimeter because of the following reasons. The total photon flux should be distributed over the area approximately equal to S = 4 $m^2$. Then the number of photons in one cubic centimeter $n_υ$ can be found from the relation:

$$N_υ = n_υ * S * c, \qquad (5)$$

where $N_υ$ = 6 * $10^{15}$ is the number of photons per second, produced on a heavy target by the microtron. From (5) we find that for the selected parameters $n_υ$ = 0.5, that there is one photon per two centimeters.

Now you can find the number of free electrons which left the smog particle under the action of the incident photons. This number for smog particle with a diameter of 10 μ, having a cross-section of $S_p$ ~ $10^{-6}$ $cm^2$ is:

$$N_e = n_υ * S_p * c * p_υ = 0.5*10^{-6} *3*10^{10}*2*10^{-5} = 0.3 \ s^{-1}. \qquad (6)$$

This means that if the smog particle remains in the field of radiation longer than 3 seconds, the probability of its ionization by the flux of the photons produced by the microtron becomes greater than 1. It means that this particle will be ionized by this photon flux.

**5. Electrostatic filter**

The electrostatic filter is a sequence of oppositely electrically charged plates, through which the air containing the electrically charged smog particles passes. Let the distance between the plates be equal to $Δ x_1$ = 1 cm and the potential difference between adjacent plates is U = 1 kV. Then, the electric field in the gap between the plates is equal to E = 1 kV / cm. When the smog particle diameter d = 10 μ, its volume is equal to: V = (4/3) π $(d / 2)^3$ = 5 * $10^{-10}$ $cm^3$. If the density of smog particle substance ρ = 2 g / $cm^3$, then the smog particle mass is: M = ρ * V = $10^{-9}$ g. Let us find the value of the extra electric charge per nucleon in the smog particle from the following considerations. Suppose the molecular weight of the substance contained in the smog particles is equal to 30 g. Let us write the following proportion:

$$6*10^{23} \text{ ----------- } 30 \text{ g}$$
$$x \text{ ------------ } 10^{-9} \text{g}, \qquad (7)$$

where $6 * 10^{23}$ - Avogadro's number. From (7) we find that the smog particle contains $6 * 10^{23} * 10^{-9} / 30 = 2 * 10^{13}$ molecules or $6 * 10^{14}$ nucleons.

Then, the amount of extra electric charge per nucleon at the single ionization of the smog particle is as follows:

$$Z/A = 1/(6*10^{14}) = 1.7*10^{-15}. \qquad (8)$$

Let the height of the vertical plates of the electrostatic filter be equal to 1 m, and the vertical velocity of the air containing the smog particles is $v_0 = 10$ cm / s. Then the air passes through the electrostatic filter during a period of time: $\tau = 1\text{m} / (10 \text{ cm} / \text{s}) = 10$ s.

We find the transverse velocity of the smog particle under the influence of the electric field from the following relationship:

$$Mdv/dt = eE. \qquad (9)$$

From the above

$$\beta = v/c = (Z/A)eEc\tau/mc^2, \qquad (10)$$

where $\beta$ is the horizontal velocity of the smog particles expressed in terms of the light velocity in vacuum, $mc^2 = 1$ GeV - the nucleon mass, expressed in energy units. The value of the transverse displacement is as follows:

$$\Delta x_2 = \beta c\tau/2 = (Z/A)eEc^2\tau^2/2mc^2 = 1.7*10^{-15}*10^3*9*10^{22}/2*10^9 = 0.7 \text{ m}. \qquad (11)$$

It can be seen that this distance $\Delta x_2 = 0.7$ m is much larger than the distance between the plates, $\Delta x_1 = 1$ cm. This means that one-time ionized smog particle cannot pass through the electrostatic filter and will stick to one of the filter plates.

Of course, from time to time the dust - smog particles, should be removed from the plates of the electrostatic capacitor. It should be removed approximately the same way as it is done from the dust bag in the vacuum cleaner.

**6. Protection against the accelerator radiation**

Although it is assumed that the microtron is placed high above the ground, on the chimney tube shown in Fig. 1, and its radiation on the ground surface must be below the maximum allowable.

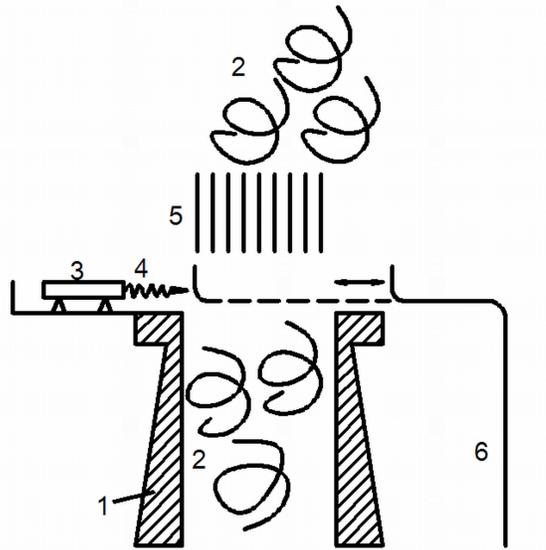

Fig.1. (1) - the boiler tube, (2) - the smoke containing particles of smog, (3) - microtron, (4) - gamma radiation, (5) - the electrostatic filter, (6) – the sleeve for transporting particles of smog from the filter to the dust collector.

Microtron produces two types of radiation. First of all, this is the gamma radiation produced in the result of interaction of the accelerated electrons with the target. To loosen the gamma quanta flux with the energy of 1MeV by $10^3$ times, it will be required to use 10 cm of lead [3], p. 964.

Neutrons will be produced on a heavy target, as a result of photonuclear reactions [4].

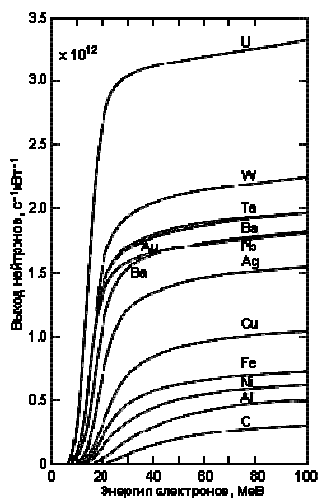

Fig.2. The intensity of the neutron flux generated on various targets, [4]

In this case for the energy of the electron beam of 25 MeV and the beam current power P = 2.5 kW approximately $4 * 10^{12}$ neutrons per second will be produced on a tungsten target. Protection against the neutrons is the hydrogen-containing substances: water, paraffin, polyethylene, and etc.

**7. Conclusion**

From the given above, it is clear that the microtron with fairly modest parameters: electron energy Є = 25 MeV and beam current I = 100 μA, will be efficiently ionize particles of smog rising together with the smoke through the boiler tube. The ionized smog particles may be removed from the smoke by means of electrostatic vertical filter plates 1 m long and electrostatic field between the plates equal to E = 1 kV / cm. The dust remained on the plates of the electrostatic capacitor should be removed into a specialized dust collector.

References

1. http://home.cern

2. http://nuclphys.sinp.msu.ru/experiment/gamma/index.html

3. Tables of physical quantities, Handbook ed. I. K. Kikoin, Moscow, Atomizdat, 1976

4. http://nuclphys.sinp.msu.ru/experiment/neutr_gen/index.html